\documentclass[12pt]{iopart}
\usepackage{graphicx}
\usepackage{epsf}
\begin{document}

\catcode`\@=11
\def\lesssim{\mathrel{\mathpalette\vereq<}}
\def\gtrsim{\mathrel{\mathpalette\vereq>}}
\def\vereq#1#2{\lower3pt\vbox{\baselineskip0pt \lineskip0pt
\ialign{$\m@th#1\hfill##\hfill$\crcr#2\crcr\sim\crcr}}}
\catcode`\@=12

\def\alt{\lesssim}
\def\agt{\gtrsim}

\vspace{2cm}

\article[CP violation]{}{
\vspace{2cm}
\begin{center}
CP violation beyond the Standard
Model\footnote[7]{Talk presented at the Proceedings of BSM Phenomenology
Workshop, May 6-11th 2001, Durham, U.K.}
\end{center}
}

\vspace{2cm}

\author{ G L Kane\footnote{gkane@umich.edu}
\dag, D D Doyle\ddag} \address{\dag\ Randall Physics
Laboratory, University of Michigan, Ann Arbor, MI 48109, USA}
\address{\ddag\ CPES, University of Sussex, Falmer, Brighton BN1 9QJ,
U.K.}

\begin{abstract}
\begin{center}
In this talk a number of broad issues are raised about the origins of
CP violation and how to test the ideas.
\end{center}
\end{abstract}
\maketitle

\section{Introduction}

The fundamental sources of CP violation in theories of physics beyond
the Standard Model is an important issue which has not been
sufficiently studied. In this talk, we begin by discussing the
possible origins of CP violation in string theory and the potential
influence
of its phenomenology on string theory. This naturally
develops into a consideration of supersymmetry breaking, specifically the
soft-breaking supersymmetric Lagrangian, $\cal{L}_{\rm
soft}$. There exist two extreme scenarios which may realistically
accommodate CP violation; the first involves small soft phases and a
large CKM phase, $\delta_{\rm CKM}$, whereas the second contains large
soft phases and small $\delta_{\rm CKM}$. We argue that there is
reasonable motivation for $\delta_{\rm CKM}$ to be almost zero and
that the latter scenario should be taken seriously. Consequently, it
is appropriate to consider what CP violating mechanisms could allow
for large soft phases, and hence measurements of these soft phases
(which can be deduced from collider results, mixings and decays,
 experiments exploring the Higgs sector or from electric dipole
moment values) provide some interesting phenomenological implications
for physics beyond the Standard Model.

\section{Fundamental properties of CP}

CP symmetry has a feature that we believe makes it an important tool for
testing ideas about fundamental theories: it is broken in nature and
yet quite general arguments indicate that it is an underlying symmetry
of string theory.  As this is a central motivation for studies of CP
beyond the standard model, it is worth first outlining these arguments
and also our hopes for how the phenomenological study of CP violation
may eventually impact on string theory.

Early studies of CP violation in string theory predominantly focused
on perturbative heterotic strings. In 1985, Strominger and Witten
showed \cite{witten} that, in string perturbation theory, CP existed
as a good symmetry that could be spontaneously broken. They argued
that a suitable extension of the four-dimensional CP operator should
reverse the direction of three of the six real compactified
dimensions; or, equivalently, that it should complex conjugate the
three complex dimensions $Z_{i}$ of the Calabi-Yau manifold. CP
violation then arises if the manifold is not invariant under the
transformation $Z_{i}\rightarrow Z_{i}^{*}$.  Alternatively (or
possibly equivalently)
it was argued that CP violation could result from CP non-invariant
compactification boundary conditions or that the breaking of CP could
be understood at the field theory level through the complex vacuum
expectation values (vev's) of moduli.


This observation was refined by Dine \etal \cite{dine} who argued
that CP is a gauge symmetry in string theory. Consequently, in
string theory, CP cannot be explicitly broken either perturbatively or
non-perturbatively. Thus string theory is a perfect example of a
theory in which all CP violating quantities (such as the ``bare
$\theta$'' of QCD, the CKM phase, $\delta_{\rm CKM}$ and the phases of
the soft supersymmetry-breaking Lagrangian, $\cal{L}_{\rm soft}^{\rm
susy}$) are enforced to be initially zero by a gauge symmetry, and all
observable CP violation arises from a spontaneous symmetry breakdown,
and is therefore calculable\footnote{An important point is that
complex phases may not lead to physically observable CP
violation. This is a well know aspect of spontaneous CP violation in
field theory, but it applies equally to string theories as discussed
by Dent \cite{dent}. In heterotic string models, CP is preserved by
any (generally complex) moduli vev's that lie on the boundary of the
fundamental domain.}.  This CP violation would then feed through to
Yukawa couplings in the superpotential {\it W} and, almost inevitably,
to phases in the soft supersymmetry-breaking terms $\cal{L}_{\rm
soft}$, which could be determined by experiment.

This has significant phenomenological implications --- if the phases in
{\it W}/$\cal{L}_{\rm soft}$ can be determined by experiment, it may
be possible to extract direct information about the patterns of the
underlying theory. Some possible revelations are the moduli dependence
of the Yukawas, potential mechanisms for supersymmetry breaking and
its transmission to the physical world, the complex dilaton (whose
imaginary part can also act as a source of CP violation) and moduli,
and the geometry of the compactification manifold.

Recent progress has led to a number of alternative approaches which
could significantly alter our understanding of CP violation. These
include Type I  or Type IIB theories with D-Branes, CP violation in
Brane worlds or, possibly, theories involving warped
compactifications. There has been some study of CP violation within these,
for example \cite{cpstring}, but further work is necessary.

We complete this section by making a more general remark about the
relationship between string theory and experiment.  It is often said
that string theory is too young a subject to be applied to the real
world, and that one should wait until it is fully developed. We would
argue precisely the opposite; in our view the {\it only} way to
develop string theory properly (that is in a direction that might have
something to do with nature) is to deduce from experimental data how
to formulate it.  Indeed, if we have learned anything from recent
progress in string theory, it is that heading in the direction marked
`more fundamental' usually reveals new ways to construct the Standard
Model, and seldom eliminates any.  Our hope for CP is therefore that,
once the CP violating parameters of $\cal{L}_{\rm soft}$ are measured
and translated to the unification scale (admittedly a difficult task),
they will aid string theorists in understanding how the extra
dimensions are compactified and supersymmetry is broken.

\section{The Soft-breaking Supersymmetric Lagrangian}\label{lag}

It is known that supersymmetry has to be a broken symmetry due to
the fact that none of the superpartners of the Standard Models
have as yet been discovered (if it were conserved, selectrons
would have masses equal to $m_{e}=0.511\;\rm Mev$, and the gluinos
and photinos would be massless). Although it is known that
supersymmetry must be broken in the vacuum state chosen by nature,
the physics of its breaking is not yet understood. There are
several potential mechanisms; it is not yet known which, if any,
are correct and neither is it known how the breaking is
transmitted to the superpartners. Despite this ignorance it is
possible to write a general, gauge invariant, Lorentz invariant,
effective Lagrangian (as discussed in \cite{georgi}). The
Lagrangian is defined to include all allowed terms that do not
introduce any quadratic divergences and depends on the assumed
gauge group and particle content.

It is expected that a realistic phenomenological model should have a
Lagrangian density which is invariant under supersymmetry, but a
vacuum state which is not. That is supersymmetry should be an exact
symmetry which is {\it spontaneously} broken. This enables
supersymmetry to be hidden at low energies in much the same way as
electroweak symmetry is concealed at low energies in the Standard
Model. A general way to describe this is to introduce extra terms in
the theory's effective Lagrangian which break supersymmetry explicitly. The
extra supersymmetry-breaking couplings should be ``soft'' (that is, of
positive mass dimension) in order for broken supersymmetry to provide
a solution to the problem of maintaining a hierarchy between the
electroweak scale and the Planck mass scale.

The effective Lagrangian for the theory can then be written in the
form

\begin{equation}\label{effectivel}
\cal L = L_{\rm SUSY}+L_{\rm soft}
\end {equation}
where the first term preserves supersymmetry invariance and the second
violates supersymmetry, (using the notation of \cite{martin})

\begin{equation}\label{lsoft}
\begin{array}{ll}
-{\cal L_{\rm
soft}}=&\frac{1}{2}\left(M_{3}\tilde{g}\tilde{g}+M_{2}\tilde{W}\tilde{W}+M_{1}\tilde{B}\tilde{B}+\rm
h.c.\right)\\
&+M_{\tilde{Q}}^{2}\tilde{Q}^{*}\tilde{Q}+M_{\tilde{U}}^{2}\tilde{U}^{c*}\tilde{U}^{c}+M_{\tilde{D}}^{2}\tilde{D}^{c*}\tilde{D}^{c}+M_{\tilde{L}}^{2}\tilde{L}^{*}\tilde{L}+M_{\tilde{E}}^{2}\tilde{E}^{c*}\tilde{E}^{c}\\
&+a_{U}\tilde{U}^{c}\tilde{Q}H_{U}+a_{D}\tilde{D}^{c}\tilde{Q}H_{D}+a_{E}\tilde{E}^{c}\tilde{L}H_{D}+
\rm h.c.\\ &+m_{H_{U}}^{2}H_{U}^{*}H_{U}+m_{H_{D}}^{2}H_{D}^{*}H_{D}+
\left(BH_{U}H_{D}+\rm h.c.\right)
\end{array}
\end{equation}

Supersymmetry is broken because these terms contribute explicitly to
masses and interactions of, for example, winos and squarks, but {\it
not} to their superpartners. The mechanisms for how supersymmetry
breaking is transmitted to the superpartners, and their interactions,
are encoded in the parameters of \eref{lsoft}.

Focusing on the phases of the soft parameters is justified for
several reasons. If they are large they can have substantial effects
on a variety of phenomena. Soft phases are a leading candidate to
explain the baryon asymmetry in the universe (the inability of the CKM
phase to achieve this is a primary reason to explore physics beyond
the Standard Model) and it has been suggested \cite{frere} that all CP
violation could arise from them. Such a scenario could be examined by
imagining $\delta_{\rm CKM}$ (which arises from the
supersymmetry-conserving superpotential if the Yukawa couplings have a
relative phase) to be very small, while allowing phases in
$\cal{L}_{\rm soft}$ (which arise from supersymmetry-breaking) to be
significant. That is, CP violation would arise {\it only} in soft
supersymmetry-breaking terms, with the CKM matrix being entirely
real. Large soft phases could also effect the relic density and
detectability of cold dark matter and rare decays. The patterns of
these phases and whether they are measured to be large or small should
reveal information on the mechanisms for breaking supersymmetry and
string compactification.

Here, the Minimal Supersymmetric Standard Model (MSSM) is assumed
to be the framework for a model of physics beyond the Standard
Model (SM). This theory is the most economical low-energy
supersymmetric extension of the SM and consists of the SM
particles and superpartners, the SM SU(3) $\times$ SU(2) $\times$
U(1) gauge group, two Higgs doublets (necessary in supersymmetry
to give masses to both the up-type quarks and to the down-type
quarks and charged leptons) and has a conserved R-parity. This
gives a total of a hundred and twenty four parameters, including
masses, flavour rotation angles and phases, which all have to be
measured (unless a compelling theory determines them).

Six of the parameters arise due to gaugino mass terms of the form
$M_{i}=\left|M_{i}\right|e^{i\phi_{i}}$. The squark and slepton
masses are in principle 3 $\times$ 3 hermitian matrices with
complex matrix elements, contributing $5\times6\times2=60$
parameters. Trilinear couplings between the sfermions and Higgs
bosons are arbitrary 3 $\times$ 3 complex matrices which
constitute $2\times9\times2=36$ parameters. Additional parameters
arise due to the gravitino, which has a mass and a phase which may
be observable in principle if it is the lightest supersymmetric
particle (LSP), and a complex effective $\mu$ term must also be
generated which is the supersymmetric version of the Higgs boson
mass in the SM and has a magnitude of the order of the other soft
terms. The symmetries of the theory allow some of these parameters
to be absorbed or rotated away by field redefinitions; in this
case, resulting in thirty-three mass eigenstates, forty-three
phases and the CKM angle.

In fact, the correct theory could be larger than the MSSM. For
example, one could want to extend the theory with an extra singlet
scalar or an additional U(1) symmetry by adding the associated
terms. To include neutrino masses in the theory, one would have to add
new fields such as right-handed neutrinos and their superpartners and
the associated terms in $\cal L_{\rm soft}$. The physics of the above
parameters is understood and is observable in many ways, so any extra
variables could be checked experimentally.

It is interesting to examine how phases will enter the theory. The
physics is embedded in the soft-breaking Lagrangian and the superpotential,
which is of the form

\begin{equation}
W\sim Y_{\alpha\beta\gamma}\phi_{\alpha}\phi_{\beta}\phi_{\gamma}
\end{equation}
where the $Y_{\alpha\beta\gamma}$ are Yukawa couplings in the scalar
field basis.  The trilinear soft terms are of the form

\begin{equation}
A_{\alpha \beta \gamma} = F^m \bigl[ \hat K_m  + \partial_m \log
Y_{\alpha \beta \gamma}-\partial_m \log (\tilde K_{\alpha} \tilde
K_{\beta} \tilde K_{\gamma} ) \bigr] \;. \label{A-terms}
\end{equation}
Here the Latin indices refer to the hidden sector fields while the
Greek indices refer to the observable fields; $F^m$ is the hidden
auxiliary fields. The K\"ahler potential is expanded in observable
fields as $K=\hat K + \tilde K_{\alpha} \vert C^{\alpha} \vert^2
+...$ and $\hat K_m \equiv \partial_m \hat K$. The A's depend on
linear combinations of Yukawas and their derivatives.  The CKM
phase arises as the relative phases of the Yukawas.  Note that if
the Yukawas have large phases it is very likely the trilinears
also have large phases, but the converse is not necessarily true.

Thus the theory suggests that if $\delta_{\rm CKM}$ is large then so
are soft phases, but soft phases could be large even if $\delta_{\rm
CKM}$ is small.  Since the baryon asymmetry cannot be described by the
SM alone, some other phases are needed; presumably the soft
phases.  Consequently it is very interesting to consider the
possibility that $\delta_{\rm CKM}$ is small \cite{frere}. As we shall see,
all CP violation today can be described by the soft phases --- there is no
phenomenological evidence \cite{cpflavour} that $\delta_{\rm CKM}$
is large.  (Of course, this does not necessarily exclude the possibility that $\it{both}$
$\delta_{\rm CKM}$ and the soft phases are large.)

Let us now discuss how most of these phases affect observables and hence how
the couplings in $\cal L_{\rm soft}$ can be measured.

\section{Measuring the Phases of $\cal L_{\rm soft}$}

Experiments measure kinematical masses of superpartners, and cross
sections $\times$  branching ratios, electric dipole moments and
so forth, whilst phenomenologists need to examine how these
measurements can be expressed in terms of the soft parameters. At
most, two soft parameters could potentially be measured directly,
the gluino mass and the gravitino mass, and the latter only if it
is the LSP and then only approximately.

Forty three of the parameters in $\cal L_{\rm soft}$ are phases
and these soft phases have effects on many observables, not just
CP violation. The measurements which are the focus of discussion
here are the EDMs of the electron, neutron and mercury; the $\it{
K}$,$\it{D}$, $\it{B}$ systems, observables being $\triangle
m_{B_{d}}, \triangle m_{B_{s}}, \epsilon, \epsilon', \sin2\beta$;
baryon asymmetry; the Higgs sector, parameters being $m_{h},
\sigma_{h}$, branching ratios and $\it{h-A}$ mixing; the relation
of superpartner masses to Lagrangian masses and the cross-sections
and branching ratios of the superpartners. Other sectors include
the decay $b \rightarrow s \gamma$, observables being its
branching ratio and CP asymmetry; rare decays such as
$K^{+}\rightarrow \pi^{+}\nu\bar{\nu}, K_{L}\rightarrow
\pi^{0}\nu\bar{\nu}$; and the relic density and detectability of
the lightest supersymmetric particle.

\subsection{Baryon Asymmetry}

It is appropriate to start by looking at the problem of baryon
asymmetry as the Standard Model cannot explain it whatever the
value of $\delta_{\rm CKM}$. There are a variety of reasonable
approaches \cite{baryons} that seek to achieve this, but in all
cases the analysis is very complicated and the resulting values
are still uncertain.

One appealing mechanism is that of Affleck-Dine baryogenesis, where
supersymmetry-breaking gives rise to a potential in the so-called
``flat directions'' (the many-parameter set of vacuum states in
supersymmetric unified theories) with a curvature of the order of
$m^{2}$ (where $m$ is the scale at which explicit soft breaking occurs
and is comparable to $M_{W}$). This small curvature allows scalar fields
to be pushed to large vev's, resulting in the Universe developing a
substantial baryon number. In this case the origin of the CP violating phase
is most likely to be supersymmetric soft phases.

A different idea involves
leptogenesis from the decay of heavy
Majorana neutrinos, or their superpartners, which have masses of the
order of $10^{11}$ GeV. In $\rm B-L$ conserving theories, sphaleron
interactions will generate a baryon asymmetry from the lepton
number-violating Majorana neutrino decay. The origin of the CP
violating phase here is rarely considered. Some possibilities are that it
occurs in the couplings of these heavy neutrinos in the
superpotential, Yukawa couplings or higher dimension operators, or in
soft terms involving Majorana fields. The mass matrix would be given
by

\begin{equation}
\left(\begin{array}{cc}m & m_{D}\\
m_{D} & M\end{array}\right)
\end{equation}
where M is related to the soft phases and $m_{D}$ contains the lepton Yukawa
phase which needs to be of a sufficient magnitude to accommodate the
level of baryon asymmetry. There are other possible mechanisms which have
been explored, such as grand unified theory (GUT) baryogenesis, which is
preserved by B-L conservation and involves GUT Yukawa phases contributing to
the
asymmetry.

A particularly attractive mechanism involves the electroweak phase
transition. It is known that $\rm B+L$ violating transitions would wash out
any net $\rm B+L$ at temperatures much higher than the weak scale. However
various processes can generate a baryon asymmetry at the electroweak phase
transition itself and, provided it is strongly first order, this asymmetry
will not be washed out by sphalerons and arises due to soft phases, not
$\delta_{\rm CKM}$.

\subsection{An example - the Chargino Mass Matrix}

An important case and the simplest example of phenomenology is the
chargino sector. The chargino mass matrix can be derived from $\cal
L_{\rm SUSY}$ and is given by (in the basis shown)
\begin{equation}\label{chgino}
M_{\tilde{C}}=\begin{array}{ccc}\begin{array}{c}\left(\begin{array}{c}M_{2}e^{i\phi_{2}}\\\sqrt{2}M_{W}\cos\beta\end{array}\right.\\\tilde{W}\end{array}&\begin{array}{c}\left.\begin{array}{c}\sqrt{2}M_{W}\sin\beta\\\mu
e^{i\phi_{\mu}}\end{array}\right)\\\tilde{H}\end{array}&\begin{array}{c}\tilde{W}\\\tilde{H}\\\;\end{array}
\end{array}
\end{equation}

When electroweak symmetry is broken and the neutral Higgs field gets
vev's, the spin $\frac{1}{2}$ fermion superpartners of the $W^{\pm}$
bosons mix with those of the charged Higgs bosons, $H^{\pm}$, producing the
above matrix. The physical mass eigenstates $M_{\tilde{C}_{1}},
M_{\tilde{C}_{2}} $ are

\begin{equation}\label{trace}
M_{\tilde{C}_{1}}^{2}+M_{\tilde{C}_{2}}^{2}=\rm{Tr}
M_{\tilde{C}}^{\dagger}M_{\tilde{C}}=M_{2}^{2}+\mu^{2}+2M_{W}^{2}
\end{equation}
\begin{equation}\label{det}
\begin{array}{ll}M_{\tilde{C}_{1}}^{2}M_{\tilde{C}_{2}}^{2}&= \rm Det
M_{\tilde{C}}^{\dagger}M_{\tilde{C}}\nonumber\\
&=M_{2}^{2}\mu^{2}+2M_{W}^{4}\sin^{2}2\beta-2M_{W}^{2}M_{2}
\mu\sin2\beta\cos(\phi_{2}+\phi_{\mu})\end{array}.
\end{equation}
with
\begin{equation}\label{tanb}
\tan\beta=\frac{\left<H_{U}\right>}{\left<H_{D}\right>}
\end{equation}

In order to relate experiment and theory, we need to measure
$\tan\beta$, $M_2$, $\mu$, $\phi_2 + \phi_\mu$.  Ultimately there are only
two mass eigenstates and four unknowns.  If more observables such as cross
sections are added, more
parameters enter such as sneutrino or squark masses.  Thus,
it is not possible, in general, to measure $\tan\beta$ and
$\mu$ and the Lagrangian parameters that include phases at hadron colliders.  Claims that
$\tan\beta$ can be ascertained at hadron colliders
are based on assumptions about soft
parameters, so these would not be direct measurements. On the other hand,
at lepton colliders
with polarized beams, and energies above the threshold for some
superpartners, it is possible to measure $\tan\beta$, $\mu$, etc.

The phases enter the masses $M_{\tilde{C}_{1}}^{2},
M_{\tilde{C}_{2}}^{2}$ in the last term of \eref{det}. Hence
it is important to note that, in general, masses (which are not CP
violating) also depend strongly on the phases.

\subsection{Phases at Colliders}

The phase of the gluino is a prime example of the subtleties of
including and measuring phases. The gluino part of the Lagrangian is
given by
\begin{equation}
{\cal L}\sim M_{3}e^{i\phi_{3}}\lambda_{\tilde{g}}\lambda_{\tilde{g}}+\rm
h.c.
\end{equation}
where $\lambda_{\tilde{g}}$ is the gluino field. This can be redefined so
that the masses are real and the vertices pick up the phases.
\begin{equation}
\psi_{\tilde{g}}=e^{i\frac{\phi_{3}}{2}}\lambda_{\tilde{g}}
\end{equation}

\begin{center}
\begin{figure}
\vspace{-3cm}
\epsfxsize=8.in
\epsffile{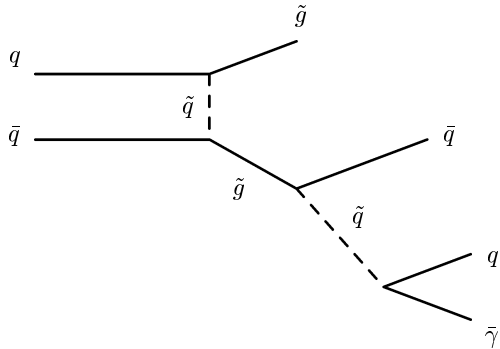}
\vspace{-20.4cm}
\caption{\label{gluino} Gluino decay}
\end{figure}
\end{center}

The Feynman rules introduce factors of
$e^{i\phi_{3}/2}$ or its complex conjugate at each of the vertices. If
the gluino then decays via a quark to (for example)
$q\bar{q}\tilde{\gamma}$, as illustrated in \fref{gluino}, a factor of $e^{i\phi_{3}/2}$ enters at the
gluino vertex and a factor of $e^{-i\phi_{1}/2}$ at the photino
vertex. This results in a differential cross section of
\begin{equation} \frac{d\sigma}{dx}\sim
m_{\tilde{g}}^{4}\left(\frac{1}{\tilde{m}_{L}^{4}}+\frac{1}{\tilde{m}_{R}^{4}}\right)\left[x-\frac{4x^{2}}{3}-\frac{2y^{2}}{3}+y\left(1-2x+y^{2}\right)\cos(\phi_{3}-\phi_{1})\right]
\end{equation}
where $x=E_{\tilde{\gamma}}/m_{\tilde{g}}$ and
$y=m_{\tilde{\gamma}}/m_{\tilde{g}}$.
$\phi_{3}$ also enters $\epsilon, \epsilon'$ in the Kaon system. See
\cite{gluino} for detailed
discussions of how various distributions depend on this phase and
other soft parameters, enabling measurements at Tevatron and LHC.

\subsection{The Higgs Sector}

\begin{figure}
\vspace{-4cm}
\epsfxsize=10.in
\hspace{-1cm}
\epsffile{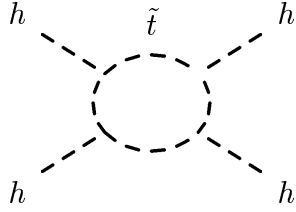}
\vspace{-27.4cm}
\caption{\label{higgs} Higgs Corrections}
\end{figure}

We now consider how the phases affect the physics of the Higgs
sector. See \cite{x} for details and further references.  The
Higgs potential includes radiative corrections of the form shown in \fref{higgs}.

The phases enter at the one loop order with stop loops being dominant
for low to medium values of $\tan\beta$. Much like the chargino mass
matrix, the stop mass matrix is in general complex, so the phases
enter into the scalar effective potential. One can write

\begin{eqnarray}\label{higgs2}
H_{d}&=&\frac{1}{\sqrt{2}}\left(\begin{array}{c}V_{d}+h_{d}+ia_{d}\\h_{d}^{-}\end{array}\right)\\
H_{u}&=&\frac{e^{i\phi}}{\sqrt{2}}\left(\begin{array}{c}h_{u}^{+}\\V_{u}+h_{u}+ia_{u}\end{array}\right)
\end{eqnarray}
with the vev's taken to be real and using \eref{tanb}. A phase $\theta =
\theta(\phi_{A_{t}},\phi_{\mu})$ allows a relative phase between the two
vev's at the minimum of the Higgs potential and cannot be rotated away. The
stop mass matrix is
\begin{equation}
m_{\tilde{t}}^{2}=\left(\begin{array}{cc}
m_{\tilde{L}}^{2}+m_{t}^{2}&Y_{t}\left(A_{t}H_{u}^{0}-\mu^{*}H_{d}^{0}\right)\\
\ast&m_{\tilde{R}}^{2}+m_{t}^{2}
\end{array}\right)
\end{equation}

The Higgs mass matrix is derived by minimizing the scalar potential,
setting $\partial V/\partial h_{d}, \partial V/\partial h_{u},
\partial V/\partial a_{d}$ and $\partial V/\partial a_{u}$ to be equal
to zero. Of the resulting four equations, only two are independent, so
three conditions remain. The Higgs sector has twelve parameters;
$V_{u}, V_{d}, \phi_{A_{t}},\phi_{\mu},\theta,|A_{t}|,|\mu|,Q$ and the $\cal
L_{\rm soft}$ parameters
$m^{2}_{\tilde{Q}},m^{2}_{\tilde{u}},b,m^{2}_{H_{u}},m^{2}_{H_{d}}$. Four
of these can be eliminated using the above three conditions and the
fact that the renormalization scale, {\it Q}, is chosen so as to
minimize any higher order corrections. Applying the conditions for
electroweak symmetry breaking (replacing $V_{u},V_{d}$ with $M_{
Z},\tan\beta$) further reduces the number of parameters to seven so
that any descriptions of the Higgs sector are now based on $\tan\beta,
\phi_{A_{t}}+\phi_{\mu}, |A_{t}|, |\mu|, m_{\tilde{L}}^{2},
m_{\tilde{R}}^{2}, b$. This number cannot be reduced without some
further theoretical or experimental information.

If $\tan\beta$ is large, sbottom loops could be large and also enter
the scalar potential, and additional parameters due to, for example,
$\tilde{C}, \tilde{N}$ loops could be significant, (see \cite{nath}). If the
phase $\phi_{A_{t}}+\phi_{\mu}$ is non-zero, it is not
possible to separate the pseudo-scalar {\it A} from {\it h, H} and it
is necessary to diagonalise a 3 x 3 matrix for neutral scalars. (In the
limit of no CP-violating phase, the three mass eigenstates $H_{i}$ are
$H_{1}\rightarrow h, H_{2}\rightarrow A, H_{3}\rightarrow H$).  These can
then decay into any given final state or could be produced in any
channel, producing three mass ($b\tilde{b}$) peaks in a decay channel
resulting in, for example, {\it Z}+Higgs. All of the branching ratios
and cross sections depend on the phase and so can change
significantly. There are two interesting phenomenological situations
to consider, depending on whether a Higgs is found at LEP or not.
\begin{center}
\begin{figure}
\epsfxsize=4.in
\hspace{2cm}
\epsffile{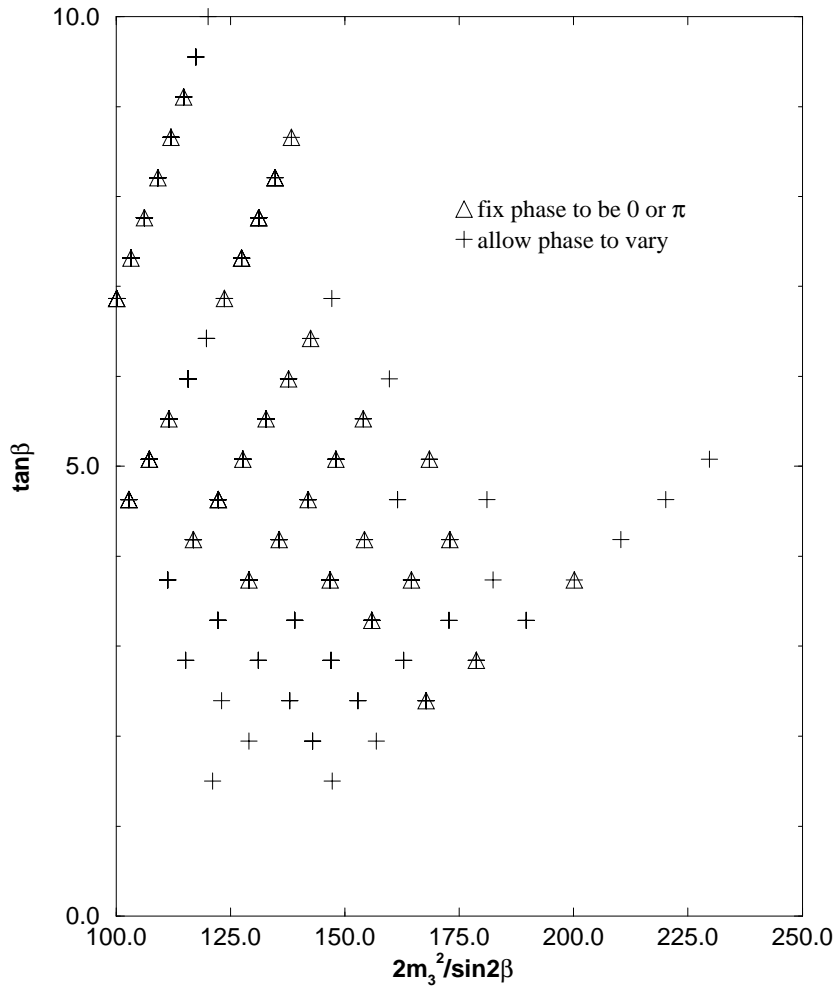}
\caption{\label{yeshiggs}Higgs found at LEP}
\end{figure}
\end{center}

\subsubsection{Higgs found at LEP or the Tevatron}

If a Higgs were found and $m_{H_{1}}$ and its $\sigma\times\rm BR$
were measured, what region of the full seven dimensional parameter
space would be allowed? There are different answers depending on whether the
phase is set to zero or $\pi$ or whether a general phase is
allowed. Consequently, it would be extremely misleading not to include
a phase in the data analysis if there were a discovery.

To illustrate this, the allowed parameter region is shown in \fref{yeshiggs} from
reference \cite{x}. The factor $2m_{3}^{2}/\sin2\beta$ on the
horizontal axis would be $m_{A}^{2}$ if there were no phase (if it was
set to zero or $\pi$). The diagram only illustrates the effect of
the phase; the full range of other parameters is not included and
experimental aspects are not taken into consideration, except for
crude estimates of the efficiencies. If the heavier Higgs were heavy
(the decoupling limit) the effect of the phase decreases for the lower
limit on the mass of the lightest eigenstate, but the effect of the
phase on the lower limit of $\tan\beta$ is still significant.
\begin{figure}
\epsfxsize=4.in
\hspace{2cm}
\epsffile{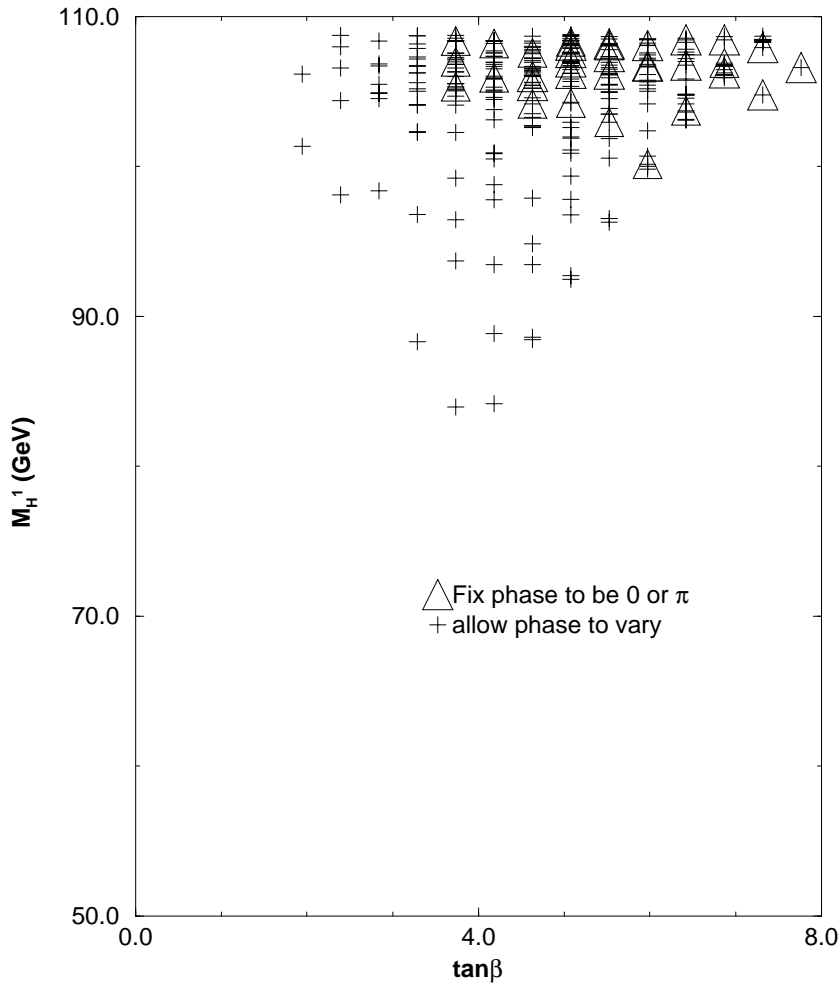}
\caption{\label{nohiggs} NO Higgs found at LEP}
\end{figure}
\subsubsection{No Higgs found at LEP}

If no Higgs were found at LEP, this would produce an experimental
limit on $\sigma(H_{1})\times\rm BR$$(H_{1}\rightarrow b\bar{b})$. So,
what would be the lower limits on $m_{H_{1}}$ and $\tan\beta$ in the
full seven parameter theory? It is clear from the diagram (\fref{nohiggs})
 that the
mass of the lightest Higgs, $H_{1}$, is allowed to be significantly
lighter if a phase is present. That is, the lower limit for the MSSM
without a phase is approximately ten percent below the SM limit,
whereas the lower limit when the phase is allowed to vary is a further
reduction of ten percent. $\tan\beta$ also has lower values allowed if the
phase is non-zero.

With seven parameters, at least seven observables would be required to
determine any of the soft parameters from the Higgs sector
alone. Potential observables include the three neutral mass
eigenstates, the charged Higgs mass, the three $\sigma \times \rm BR$
for $\rm {\it Z} + Higgs$ and the three $\sigma \times \rm BR$ for the mass
eigenstate channels, and finally the two stop mass eigenstate
masses. It could also be possible to measure the ratio

\begin{equation}
R=\frac{\sigma(gg\rightarrow H^{2}\rightarrow
b\bar{b})}{\sigma(gg\rightarrow H^{1}\rightarrow b\bar{b})}.
\end{equation}
If all of these measurements could be made $\tan\beta$ and
$\phi_{A_{t}}+\phi_{\mu}$ could be measured directly in the Higgs
sector and the former could be compared with results from the gaugino
sector. Such analysis could be enabled with LHC data, but it is unlikely.

\subsection{Electric Dipole Moments}

The most stringent constraints on models for sources of CP violation
come from the experimental upper limits on the absolute values for the
electric dipole moments (EDMs) of the electron, neutron and mercury
atom. The Standard Model predicts very small values (an upper limit of
order $10^{-32}{\it e}.\rm cm$), which are consistent with experimental
constraints (upper limits of $4.3\times10^{-27}$\cite{electron},
$6.3\times10^{-26}$( 90\% C L)\cite{neutron} and
$2.1\times10^{-28}$\cite{hg} {\it e}.cm). However generic
supersymmetry models predict much higher values, for example,
$d_{n}\sim 10^{-23}e.\rm cm$. As a result, it is necessary to suppress
the CP violating phases responsible for EDMs. There are a variety of
models suggested to achieve this \cite{edm}, including models with
small CP phases, models with heavy sfermions, cancellation scenarios
and models with flavour off-diagonal CP violation.

It has long been known that supersymmetry predicts values for the
neutron and electron EDMs that are approximately fifty times the
experimental limits if all the soft masses and phases are
independent. From this, it could be naively concluded that in a
supersymmetric world EDMs should have already been observed. However,
in any theory there can be relations among the soft phases that lead
to large cancellations which can occur over a large region of
parameter space even if all the soft phases are present and large.

There are in fact two possible methods of avoiding the constraints the
dipole moment measurements place on supersymmetric models. One is to
assume that all the supersymmetry phases are zero or unnaturally small
($\alt 0.01$). The second possibility is that the phases may be large
while certain approximate relations hold among the mass parameters and
phases, resulting in cancellations of the order of 5 - 10 in the EDM
calculations. (Note that some cancellation effects were neglected in
earlier analyses, such as $\tilde{C}-\tilde{N}$ from the
Lagrangian). However, both these possibilities seem ``unnatural''
without some deeper understanding of what is going on. No symmetry or
dynamics is known that would imply phases are small.  One can think of
arguments such as dilaton dominance, but that is an ad hoc and not well
motivated choice unless it is determined by some deeper argument.  The
second scenario could only become acceptable if the cancellations were
due to some symmetry of high scale theory, a condition that looks like
fine tuning if we can only see the low energy theory.

\section{A string-motivated model}

\begin{center}
\begin{figure}
\vspace{-3.5cm}
\epsfxsize=8.in
\epsffile{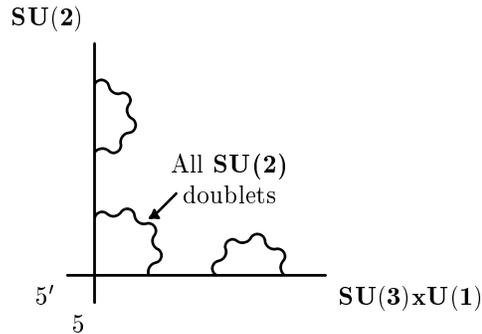}
\vspace{-20.4cm}
\caption{\label{branes}Embedding the SM on two intersecting 5-branes}
\end{figure}
\end{center}
It is now important to return to the general structure of $\cal L_{\rm
soft}$
and consider potential
models which will reduce the number of soft parameters. String models
can provide some motivation for large phases in the soft breaking
parameters, suggesting that low energy data revealing the patterns of
cancellations could reveal clues about high (Planck) scale theory. A
particularly interesting class results from embedding the SM on
D-branes within simple Type II B models \cite{dbrane}.  Open Type I
strings on D-branes that intersect at some non-vanishing angle give
rise to chiral fermions and explicit compactifications with
intersecting branes exist. An acceptable phenomenological example
arises from assuming supersymmetry breaking effects are communicated
dominantly via the F-component vev's of the dilaton and moduli \cite{munoz}.
Consider, for example, the case of two intersecting 5-branes shown in \fref{branes}.

The SM gauge group is chosen to be embedded in such a way that SU(3)
and U(1) are associated with one of the intersecting branes, and SU(2)
is associated with the other.  We only note here the resulting phase
structure

\begin{equation}
\begin{array}{rl}
m_{1}=m_{3}&=-A_{t}\sim e^{-i\alpha_{1}}\\ m_{2}&\sim e^{-i\alpha_{2}}
\end{array}
\end{equation}
and all the other soft terms are real. Then using symmetries and
rotations, the previous one hundred and five parameters can be reduced
to just eight, $\alpha_{2}-\alpha_{1},$ the mass scale
$m_{3/2},\;\tan\beta,\;|\mu|$ (from electroweak symmetry breaking),
$\phi_{\mu},$ and the relative amounts of dilaton and
moduli, $X_{1},\;X_{2}\;\rm and\; X_{3}$. It is not yet understood
fully how to include a reliable mechanism that provides the effective
$\mu$ parameter in models such as these so, for now, we treat it as an
arbitrary complex parameter. With this model it is then possible,
qualitatively, to  describe all CP violation with no contributions from the
CKM phase $\delta_{\rm CKM}$, by using a ``different'' flavour
structure \cite{cpflavour}.

\section{Fine tuning?}

The gluino sector of the above model provides an example of the fact that although some
relations may appear to be fine tuned in a low energy theory, they can
originate in the structure of the high energy theory. If the
gluino-squark box diagram, which would probably be one of the dominant
contributions to CP violation in the {\it K} system, is to be
consistent with observed values of $\epsilon$ and $\epsilon'$, the
argument of the gluino phase must satisfy (see \cite{finetune})

\begin{equation}\label{arg}
\rm arg\left(\left(\delta_{12}\right)_{LR}^{d}M_{3}^{*}\right)\approx
10^{-2}
\end{equation}
where $\left(\delta_{12}\right)_{LR}^{d}\sim A_{12}^{d}$. This value can be
achieved in models with large flavour violation in the A-terms. It seems
like fine tuning, but in a D-brane model, assuming that
\begin{equation}
\phi_{\left(\delta_{12}\right)_{LR}^{d}}=\phi_{A_{sd}}=\phi_{M_{3}}
\end{equation}
would force $\rm arg\left(\left(\delta_{12}\right)_{LR}^{d}M_{3}^{*}\right)$
to be zero at high energies. However, the phases run differently and can
generate the factor of $10^{-2}$.

\section{The {\it K} and {\it B} systems}

The angle $\beta$ is one of the angles of the CKM unitarity triangle
and is an important test for physics beyond the SM.
The SM ratio $|V_{ub}/V_{cb}|$ is generally thought of to be unaffected by
BSM physics as it arises from tree level decays, but all the
other constraints ($\epsilon_{K}$, $\triangle m_{B_{d}}$ and
$\triangle m_{B_{s}})$ can be affected. It is possible
that the unitarity triangle could be flat, giving a $\beta$ and hence
$\sin2\beta$ of zero for the SM where $\sin2\beta$ is a measure of the CP
asymmetry for the decay $B^{0}\bar{B}^{0}\rightarrow \psi
K_{K,L}$; a large $\sin2\beta$ could arise from soft phases.

\begin{center}
\begin{figure}
\vspace{-4.5cm}
\epsfxsize=8.in
\hspace{-3.7cm}
\epsffile{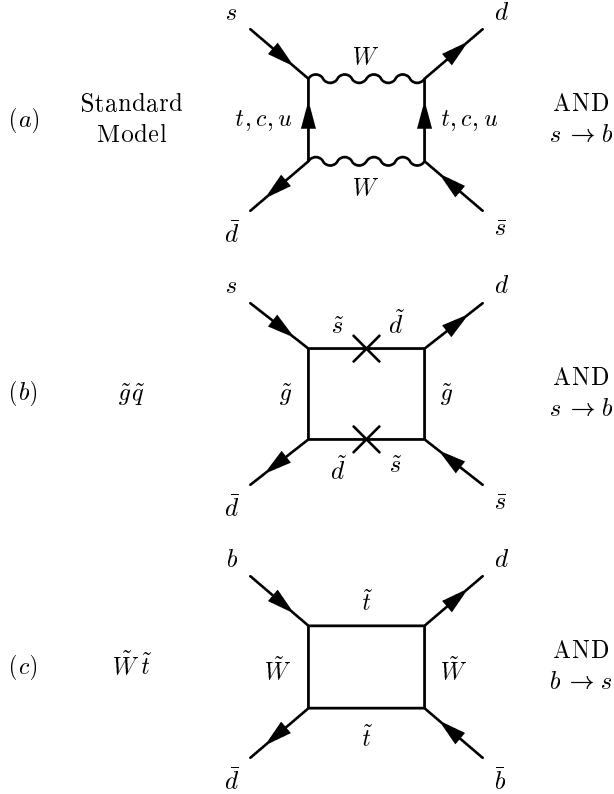}
\vspace{-14cm}
\caption{\label{Bsys} Loop contributions in the K and B systems}
\end{figure}
\end{center}

The Feynman diagrams in \fref{Bsys} show the loop contributions to mixing
(there are analogous penguin diagrams for decays contributing to
direct CP violation).
For both the {\it K} and {\it B} systems, the supersymmetry effects
arise from loops. As mentioned in the previous section, within the
{\it K} system the dominant contributions are most likely to be
gluino-squark boxes (and penguins). The first diagram is the usual SM
box diagram for calculating the value of the neutral Kaon mixing
parameter $\triangle m_{K}$, but supersymmetric boxes would also
contribute. With the postulate that $\delta_{\rm CKM}\approx 0$, the
SM box would be entirely real and would not contribute to
$\epsilon_{K}$, the indirect CP violating parameter. The
supersymmetric box shown in the second diagram is the gluino-squark
box discussed in the previous section (the ``x'' refers to a L-R
chirality flip). The magnitude of $A_{sd}$, the triscalar coupling,
must be of the right size to describe $\epsilon_{K}$ and the
supersymmetric approach is only able to describe it, not explain
it. In this case, in fact, $\epsilon_{K}$ is described more naturally
by the SM.

Within the B system, all decays (except $b\rightarrow s\gamma$) have a
tree level contribution, implying that the B system with $\delta_{\rm
CKM}\approx 0$ is superweak and all CP violating effects
arise due to mixing, with $\epsilon_{B}' \approx 0$. The dominant
mixing is usually assumed to be caused by the chargino-stop box shown
in the third diagram. It is predicted from this that because the decay
phase is zero,
\begin{equation}
\sin2\alpha\; =\; -\sin2\beta
\end{equation}
where $\sin2\alpha$ and $\sin2\beta$ are defined to be the CP
asymmetries measured in the decays $B_{d}\rightarrow \pi^{+}\pi^{-}$
and $\psi K_{s}$. It is also known that if $\delta_{\rm CKM}\approx
0$, then we can take

\begin{equation}
V_{td}=|V_{cb}\sin \theta_{c}|-|V_{ub}|\approx 0.005
\end{equation}
These results do not depend on the soft parameters and so act as an
independent test of the approach. Studies show that there are regions
of parameter space where the values for recent measurements and the
neutral {\it B} mixing parameter can be achieved, using the estimate
for $\sin2\beta$ calculated by this model.

\section{Some further notes}

The recent measurement of $g_{\mu}-2$ and the LEP Higgs lower limit
suggest that $\tan\beta \agt 5$. The natural value is of the order
$\tan\beta \sim 1$, the supersymmetric limit, and a naive estimate
using Yukawa unification is $\sim 35$.

CP violation in the lepton sector is very interesting.  It will also arise
via
the superpotential Yukawa matrix (for leptons) and soft phases.  We
will not discuss it further in this talk.

The decay $b\rightarrow s\gamma$ has a CP asymmetry which does not
involve any tree level contribution. The SM predicts a value of the
order of half a percent, whereas supersymmetric estimates range up to
fifteen percent or more. This difference is interesting as the decay
is a relatively clean one and could potentially be the first place
that physics beyond the SM is found.

An important consideration is whether the CPT theorem is in fact only
approximate and CPT symmetry could be minimally violated. As yet there
is no good theoretical motivation for this to occur.

In \sref{lag} it was shown that the MSSM had forty three soft phases,
but a question remains as to which phases are constrained by which
experiments. For example, the measurement of $g_{\mu}-2$ constrains
$\phi_{2}+\phi_{\mu}$ if $\tan\beta$ is large and it constrains
$\phi_{2}+\phi_{\mu}$, $\phi_{1}+\phi_{\mu}$ and
$\phi_{A_{\mu}}+\phi_{\mu}$ in general. The decay $b\rightarrow
s\gamma$ constrains $\phi_{A_{t}}+\phi_{\mu}$, $\phi_{2}+\phi_{\mu}$
and $\phi_{3}+\phi_{\mu}$, whereas the Higgs sector only constrains
$\phi_{A_{t}}+\phi_{\mu}$.  Sometimes insufficient care is taken to
ensure that one is examining the relevant reparameterization invariant
phases.

\section{Outlook}

It has been argued that a softly broken supersymmetric Lagrangian can
provide a suitable framework to accommodate CP violation effects. This
involved the requirement that some soft phases were large, which would
need to be supported experimentally. Experimental results that could
achieve this include the observation of an electron EDM, the ratio of
neutron to mercury EDMs being different from what would occur with
only a strong CP phase, a value of $\sin2\beta$ which is not equal to
that predicted by the SM, Higgs sector observations, measurements of
superpartner masses and $\sigma\times \rm BR$ at colliders and the
decay $K_{L}\rightarrow \pi^{0}\nu\bar{\nu}$ not being equal to that
predicted by the SM. On the other hand evidence for non-zero
$\delta_{\rm CKM}$ could possibly be observed at b-factories or in the
decay $K_{L}\rightarrow \pi^{0}\nu\bar{\nu}$. Optimistically these
issues could be clarified over the next years at BaBar, BELLE, and
CDF, by ever more stringent limits on the values (and perhaps even
discovery) of EDMs and confirmation of the result for $g_{\mu}-2$.


\section*{References}
\begin{thebibliography}{2}
\bibitem{witten} Strominger A and Witten E 1985 {\it
Commun.Math.Phys.}{\bf101} 341
\bibitem{dine} Dine M, Leigh R G and MacIntire D 1993 hep-ph/9307152
\bibitem{dent} Dent T 2000 hep-th/0011294
\bibitem{cpstring} Abel S A and Servant G 2001 hep-ph/0105262
\bibitem{georgi} Dimopolous S and Georgi H 1981 {\it \NP}{\bf B193} 150
\nonum Girardello L and Grisaru M 1982 {\it \NP}{\bf B194} 65
\bibitem{martin} Martin S P 1999 hep-ph/9709356
\bibitem{frere} Frere J M and Gavela M B 1983 {\it \PL}{\bf B132}
\nonum Lim C S 1991 {\it \PL}{\bf B256} 233
\nonum Choi K, Kaplan D and Nelson A 1993 {\it \NP}{\bf B391} 515
\nonum Kobayashi T and Lim C S 1995 {\it \PL}{\bf B343} 122
\nonum Frere J M and Abel S A 1997 {\it \PR}{\bf D55} 1623
\bibitem{baryons} Dolgov A 1992 {\it \PR}{\bf 222C} 309
\nonum Engvist K 2000 hep-ph/0002125
\nonum Barbieri R, Creminelli P, Strunia A and Tetradis N 1999
hep-ph/9911315
\nonum Riotto A and Trodden M 1999 {\it Ann. Rev. Nucl. Sci} {\bf 49} 35,
hep-ph/9901362
\nonum Brhlik M, Good G and Kane G L 1999 hep-ph 9911243
\bibitem{tanref} Brhlik M and Kane G L 1998 {\it \PL}{\bf B437} 331
\bibitem{gluino} Mrenna S, Kane G L and Wang L-T 1999 hep-ph/9910477
\bibitem{nath} Ibrahim T and Nath P 2000 hep-ph/0008237
\bibitem{electron} Commins E D \etal 1994 {\it \PR}{\bf A50} 2960
\bibitem{neutron} Harris P G \etal 1999 {\it \PRL}{\bf 82} 904
\bibitem{hg} Romalis M V, Griffith W C anf Forston E N 2001 {\it \PRL}{\bf
86} 2505
\bibitem{edm} Brhlik M, Good G J and Kane G L 1998 hep-ph/9810457
\nonum Pokorski S, Rosiek J and Savoy C A  1999 hep-ph/9906206
\nonum Ibrahim T and Nath P 1998 {\it \PR}{\bf D58} 111301; errata-{\it
ibid}. 1999 {\bf D60} 099902
\nonum Ibraham T and Nath P 1998 {\it \PR}{\bf D57} 478; errata-{\it ibid}.
1998 {\bf D58} 019901, 1999 {\bf D60} 079903
\nonum Falk T, Olive K A, Pospelov M and Roiban R 1999 {\it \NP}{\bf B560} 3
\nonum Abel S, Khalil S and Lebedev O 2001 hep-ph/0103320
\bibitem{dbrane} Brhlik M, Everett L, Kane G L and Lykken J 1999
hep-ph/9905215
\bibitem{munoz} Ibanez L E, Munoz C and Rigolin S 1999 hep-ph 9812397
\bibitem{cpflavour} Brhlik M, Everett L, Kane G L, King S F and Lebedev O
1999 hep-ph 9909480
\bibitem{finetune} Gabbiani F \etal 1996 {\it \NP}{\bf B477}321
\nonum Bertolini S \etal 1991 {\it \NP}{\bf B353} 591
\nonum Masiero A and Murayama H 1999 {\it \PRL}{\bf 83} 9107
\bibitem{x} G.L. Kane and L-.T. Wang, Phys. Lett. 2000 {\bf B488} 383
\end{thebibliography}
\end{document}